\documentclass{article} 
\usepackage{iclr2027_conference,times}
\usepackage{graphicx}
\usepackage{booktabs} 
\usepackage{amssymb}

\usepackage{amsmath,amsfonts,bm}

\def\eqref#1{equation~\ref{#1}}

\def\1{\bm{1}}

\DeclareMathAlphabet{\mathsfit}{\encodingdefault}{\sfdefault}{m}{sl}
\SetMathAlphabet{\mathsfit}{bold}{\encodingdefault}{\sfdefault}{bx}{n}

\usepackage{hyperref}
\usepackage{url}
\usepackage{algorithm}
\usepackage{algpseudocode}
\usepackage{graphicx}
\usepackage{xcolor}
\usepackage{subcaption}

\definecolor{RWBlue}{HTML}{5782B7}
\definecolor{RWOrange}{HTML}{E5943F}
\definecolor{RWGreen}{HTML}{6AA54E}
\definecolor{RWRed}{HTML}{BE5043}

\newcommand{\RWEvalLegend}[4]{%
    \makebox[\linewidth][c]{{\scriptsize\ttfamily
    \textcolor{RWBlue}{\rule{0.7em}{0.6em}}\hspace{0.3em}#1\hspace{0.7em}%
    \textcolor{RWOrange}{\rule{0.7em}{0.6em}}\hspace{0.3em}#2\hspace{0.7em}%
    \textcolor{RWGreen}{\rule{0.7em}{0.6em}}\hspace{0.3em}#3\hspace{0.7em}%
    \textcolor{RWRed}{\rule{0.7em}{0.6em}}\hspace{0.3em}#4}}%
    \par\vspace{2pt}%
}

\newcommand{\RWEvalSubcap}[1]{%
    \par\vspace{2pt}{\centering\footnotesize\fontfamily{phv}\selectfont #1\par}%
}

\newcommand{\RWEvalPairCap}[5]{%
    \begin{subfigure}[t]{0.495\linewidth}
        \centering
        \includegraphics[page=#2,width=\linewidth]{figs/#1.pdf}%
        \RWEvalSubcap{#4}
    \end{subfigure}\hfill%
    \begin{subfigure}[t]{0.495\linewidth}
        \centering
        \includegraphics[page=#3,width=\linewidth]{figs/#1.pdf}%
        \RWEvalSubcap{#5}
    \end{subfigure}%
    \par
}

\title{RailWave: Adaptive Spatial and Temporal Scheduling for Expert-Parallel Communication}

\iclrpreprint

\renewcommand\footnotemark{}
\author{
\!Chutian Wang$^{1,*}$, Wenhao He$^{2,*}$, Jingmin Zhu$^{3}$, Qingyu Yin$^{4}$, Heng Xu$^{5,\dagger}$, Xiuyu Li$^{6,\dagger}$
\thanks{$^*$Equal contribution.}
\thanks{$^\dagger$Corresponding author.} \\
$^1$Sun Yat-sen University\quad $^{2}$Southeast University\quad $^{3}$Monash University\\
$^{4}$National University of Singapore\quad $^{5}$Shenzhen University of Advanced Technology \\
$^{6}$Renmin University of China\\
\footnotesize\texttt{\{csevanvan1024, xiuyuli0506\}@gmail.com}
}

\begin{document}

\maketitle

\begin{abstract}
Irregular All-to-All communication is a major bottleneck in expert-parallel Mixture-of-Experts (MoE) models. Even with fixed expert routing and placement, uneven utilization of parallel network Rails and incast can limit communication performance. We present RailWave, a phase-adaptive communication layer built on DeepEP that addresses these bottlenecks below the routing layer through spatial and temporal traffic shaping. RailBalance redistributes source traffic across eligible Rails using source-local information, while a reusable, topology-derived permutation schedule limits concurrent senders per receiver without rebuilding demand-dependent schedules for each communication phase. A lightweight calibrated selector chooses an execution path according to each phase's traffic characteristics and offline profiling results. On training-derived communication workloads from the 106B GLM-4.5-Air model, RailWave delivers up to $5.84\times$ speedup on H800 and $4.36\times$ on H20 over Native.
Code is available at \url{https://github.com/CyberSecurityErial/RailWave-EP}.

\end{abstract}

\section{Introduction}
\label{sec:introduction}

Mixture-of-Experts (MoE) models increase model capacity while activating
only a subset of parameters for each token
~\citep{shazeer2017outrageously,lepikhin2021gshard,fedus2022switch}.
Expert parallelism (EP) distributes experts across GPUs, requiring
dispatch and combine exchanges between tokens and their selected experts.
Because routing induces uneven and input-dependent traffic, these exchanges
form irregular All-to-All communication and can become a major bottleneck
in MoE execution~\citep{liao2025mixnet,li2023accelerating}.

Existing approaches improve MoE efficiency through routing, expert placement,
and replication
~\citep{lewis2021base,zhou2022mixture,nie2023flexmoe,he2022fastermoe}.
Replication-based methods can additionally incur per-layer memory overhead
from redundant expert states~\citep{eplb}. These mechanisms shape the logical
communication demand, but do not fully determine its physical execution. This distinction is particularly important
in Rail-optimized clusters, where a high-bandwidth intra-node fabric connects
multiple scale-out paths~\citep{nvidia2023superpod,wang2024railonly}.
Even with an optimized dispatch--combine backend such as
DeepEP~\citep{deepseek2026deepep}, traffic can still be redistributed across
eligible Rails and node-pair transfers can overlap in different orders.

\begin{figure}[t]
    \centering
    \includegraphics[trim=0 0 0 25bp,clip,width=\linewidth]{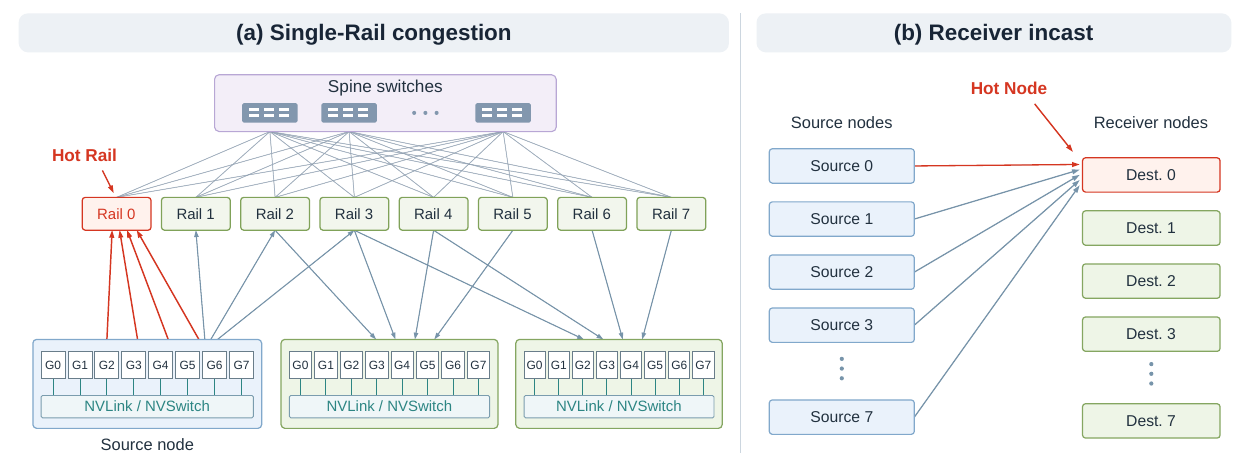}
    \par\vspace{2pt}
    \noindent\makebox[0.6\linewidth]{\small (a) Rail imbalance}%
    \makebox[0.4\linewidth]{\small (b) Receiver incast}
    \caption{Two bottlenecks with fixed routing.
    (a) Balanced expert and destination-node loads can coexist with a
    hot source Rail.
    (b) Balanced source Rails can still feed the same receiver concurrently.}
    \label{fig:motivation-fig}
\end{figure}

These execution choices expose two bottlenecks, as shown in Figure~\ref{fig:motivation-fig}: (a) Balanced expert and
destination-node loads can coexist with concentrated source-Rail traffic,
leaving parallel paths underutilized. (b) Conversely, balanced source Rails do
not prevent multiple senders from converging on the same receiver and
creating incast. Our analysis of 900
training step--layer pairs further shows that expert-load imbalance has
limited linear explanatory power for Rail imbalance. We therefore jointly control where
traffic enters the scale-out network and which node-pair transfers overlap.

We present RailWave, a communication layer built on DeepEP for spatial and
temporal control of expert-parallel traffic. RailBalance redistributes
outgoing traffic across eligible Rails using source-local information,
without requiring destination-side balancing. Reusable topology-derived
permutation scheduling organizes node-pair transfers into communication
waves that bound incast degree without rebuilding a demand-dependent
schedule for each phase.

The two mechanisms are complementary: traffic that is balanced across Rails
over an entire phase can become concentrated within an individual wave.
RailWave therefore exposes four execution paths combining these mechanisms
and uses a lightweight calibrated selector to choose among them according
to the current traffic characteristics and validated offline profiling
results.

This work makes four contributions:
\begin{itemize}
    \item We present RailWave, a DeepEP-based communication layer for joint
    control of Rail allocation and node-pair overlap.
    \item We design RailBalance, a source-local mechanism for shaping
    traffic across parallel Rails.
    \item We adapt reusable topology-derived permutation scheduling
    to bound incast degree across changing routing phases.
    \item We evaluate RailWave on two 32-GPU H800 and H20 clusters using
    real GLM-4.5-Air training workloads, achieving up to $5.84\times$ and
    $4.36\times$ communication speedups, respectively, over Native
    execution.
\end{itemize}

\section{Preliminaries}
\label{sec:prelim}

\subsection{Expert-parallel communication}
\label{sec:prelim-ep}

Under EP, \emph{dispatch} sends routed token representations to remote
experts, and \emph{combine} returns their outputs. Routing and expert
placement determine the participants and volume of these irregular
All-to-All exchanges.

We consider an EP group spanning $N$ nodes. In one dispatch or combine
operation, $D_{s,d}$ denotes the traffic volume from source node $s$ to
destination node $d$. We collect these traffic volumes in the node-level
demand matrix
{\small
\begin{equation}
\begin{aligned}
    D &= [D_{s,d}]_{s,d=1}^{N}
    \in \mathbb{R}_{\geq 0}^{N\times N}.
\end{aligned}
\label{eq:demand-matrix}
\end{equation}
}

Routing, expert placement, and token-to-replica assignment fix this demand
before execution. RailWave changes neither $D$ nor the token's destination
expert. Instead, it controls the local egress path and the concurrency of
node-pair transfers using the DeepEP backend~\citep{deepseek2026deepep}.

\subsection{Rail-optimized networks and traffic descriptors}
\label{sec:prelim-rails}

Each node contains $G$ accelerators connected by a high-bandwidth
intra-node fabric and exposes $R$ scale-out Rails. We index the Rails by
$r\in\{0,\ldots,R-1\}$~\citep{nvidia2023superpod,wang2024railonly}.
Local redistribution can move traffic to another eligible egress Rail
before inter-node transmission.

For node pair $(s,d)$, let $\chi_{s,d}(r)$ map source egress Rail $r$ to
the destination ingress Rail. We call the topology \emph{Rail-symmetric}
when this mapping is a fixed bijection for every node pair. With matching
Rail indices, the correspondence reduces to
{\small
\begin{equation}
\begin{aligned}
    \chi_{s,d}(r) &= r.
\end{aligned}
\label{eq:rail-correspondence}
\end{equation}
}

Let $C_{s,r,d}$ count tokens assigned to source egress Rail $r$ before
redistribution and routed to node $d$. Counts are aggregated according to
the GPU-to-egress-Rail mapping, which may assign multiple GPUs to one Rail.
The eligible source Rails for node pair $(s,d)$ satisfy
{\small
\begin{equation}
\begin{aligned}
    \mathcal{R}_{s,d}
    &\subseteq \{0,\ldots,R-1\}.
\end{aligned}
\label{eq:eligible-rails}
\end{equation}
}

We characterize each communication phase using three traffic descriptors
that capture its scale, source-Rail imbalance, and receiver-side
concentration. We define the average remote payload per GPU as
{\small
\begin{equation}
\begin{aligned}
    B
    &=
    \frac{\sum_{s\ne d} D_{s,d}}{NG}.
\end{aligned}
\label{eq:remote-payload}
\end{equation}
}

Let $L_{s,r}$ denote the outgoing inter-node traffic carried by Rail $r$
of source node $s$. We define the global source-Rail skew as
{\small
\begin{equation}
\begin{aligned}
    \rho_g
    &=
    \frac{\max_{s,r} L_{s,r}}
    {\frac{1}{NR}\sum_{s,r} L_{s,r}}.
\end{aligned}
\label{eq:global-rail-skew}
\end{equation}
}

A value of one indicates equal load across all source Rails.

Receiver-side concentration depends on which transfers overlap. For a
sending window $w$, let $A_w$ denote its active directed node pairs. The
traffic received by node $d$ in that window is
{\small
\begin{equation}
\begin{aligned}
    R_{w,d}
    &=
    \sum_{s:(s,d)\in A_w} D_{s,d}.
\end{aligned}
\label{eq:receiver-load}
\end{equation}
}

We then define receiver concentration as
{\small
\begin{equation}
\begin{aligned}
    \kappa
    &=
    \max_w
    \frac{\max_d R_{w,d}}
    {\frac{1}{N}\sum_d R_{w,d}}.
\end{aligned}
\label{eq:receiver-concentration}
\end{equation}
}

Thus, $B$ captures communication scale, $\rho_g$ captures source-Rail
imbalance, and $\kappa$ captures receiver concentration under the
actual sending windows.

\subsection{Communication bottlenecks with fixed routing}
\label{sec:prelim-bottlenecks}

Figure~\ref{fig:motivation-fig} illustrates the distinction: source-side
redistribution addresses Rail imbalance, whereas transfer ordering controls
incast.

\begin{figure}[t]
    \centering
    \includegraphics[width=\linewidth]{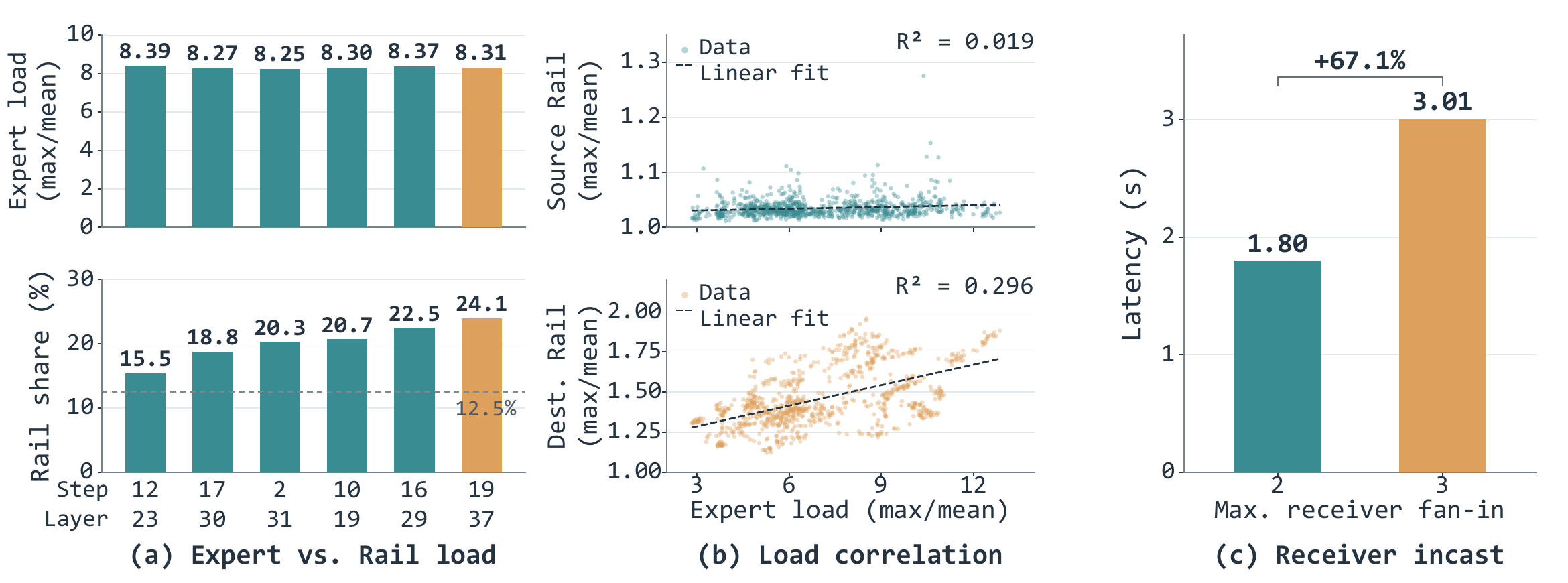}
    \caption{Rail-load imbalance and incast under fixed routing.
    (a) Six cases with similar expert-load imbalance but different
    destination-Rail concentration; the two label rows identify training
    step and layer. (b) Linear fits over 900 training step--layer pairs.
    (c) Native latency under two replay orders with identical
    rank-to-rank demand.}
    \label{fig:motivation-exp}
\end{figure}

To examine the first distinction, we analyze routing records from 900
training step--layer pairs and derive Rail loads using the expert-to-Rail
mapping. Figure~\ref{fig:motivation-exp}(a) selects six cases whose
expert-load max/mean ratios lie between $8.2$ and $8.4$. Their busiest
destination-Rail assignment shares range from $15.5\%$ to $24.1\%$.
Across all records, linear fits from expert-load imbalance to source- and
destination-Rail imbalance yield $R^2=0.019$ and $0.296$, respectively
(Figure~\ref{fig:motivation-exp}(b)). Expert-load imbalance therefore has
limited linear explanatory power for the observed Rail-load variation.

Figure~\ref{fig:motivation-exp}(c) quantifies the impact of incast on
communication latency. We keep the rank-to-rank demand, payload, backend,
and number of communication rounds fixed, while varying which node pairs
transmit concurrently. The sum of per-layer P50 dispatch--combine latencies
increases from $1.801$ to $3.009$\,s, a $67.1\%$ increase. This result
shows that the same rank-to-rank demand can incur higher communication
latency under stronger incast.

\subsection{MoE communication optimization}
\label{sec:prelim-optimization}

MoE communication has been optimized at several layers of the software
stack. Upper-layer methods change the demand presented to communication:
EPLB, UltraEP, MoonEP, and TAOT adjust expert placement, replication, or
token-to-replica assignment to balance expert or device loads
~\citep{eplb,wei2026ultraep,moonshot2026moonep,zhang2026taot}.
NetMoE adjusts input-sample placement to exploit network
locality~\citep{liu2025netmoe}, while node-limited routing restricts the
destinations selected for each token~\citep{deepseekv3}. These decisions
shape the logical demand matrix $D$. RailWave operates after them and
optimizes how the resulting communication is executed.

EP communication libraries optimize the dispatch--combine data path
across intra- and inter-node fabrics. DeepEP and NCCL-EP use high-bandwidth
intra-node communication to stage or redistribute tokens before scale-out
transmission~\citep{deepseek2026deepep,goldman2026ncclep}.
SwiftEP reduces data-path overhead through buffer fusion and TMA
offloading~\citep{li2026swiftep}, while UBEP combines finer-grained execution
with topology-aware scheduling for high-bandwidth
superpods~\citep{liu2026ubep}. EPIC spans logical and physical optimization,
coupling expert migration with topology-adaptive transport and
communication--computation overlap~\citep{cao2026balancing}.
RailWave builds on DeepEP to control the Rail distribution and concurrent
execution of the transfers carried by this backend.

Network-aware mechanisms are the closest to RailWave. RailS and FuseLink
primarily improve path utilization through local traffic redistribution
~\citep{rails,fuselink}, whereas RailWave additionally controls which
node-pair transfers overlap at a receiver. FAST explicitly balances traffic across the
GPUs and NICs within each node, then uses demand-dependent Birkhoff
scheduling to construct balanced one-to-one stages
~\citep{fast,wu2026hierarchicalbvn}. This planning requires a consistent
view of source--destination demand across nodes; synchronizing that
information and rebuilding schedules can incur substantial overhead as
routing changes. RailWave instead reuses topology-derived cyclic peer
plans with source-local RailBalance, reducing this coordination and
plan-construction overhead.
Rate-based scheduling controls incast through sender pacing~\citep{incastfree}.
SABRE selects
communication strategies according to endpoint-volume skew~\citep{sabre},
whereas RailWave indexes offline-calibrated path decisions by payload,
global source-Rail skew, and receiver concentration.

Other approaches reduce payload size or overlap communication with
computation~\citep{nie2024lsh,jiang2024lancet,zhang2025comet}. These
optimizations are complementary to RailWave's execution control.

\section{System Design}
\label{sec:system}

\begin{figure}[t]
    \centering
    \includegraphics[width=\linewidth]{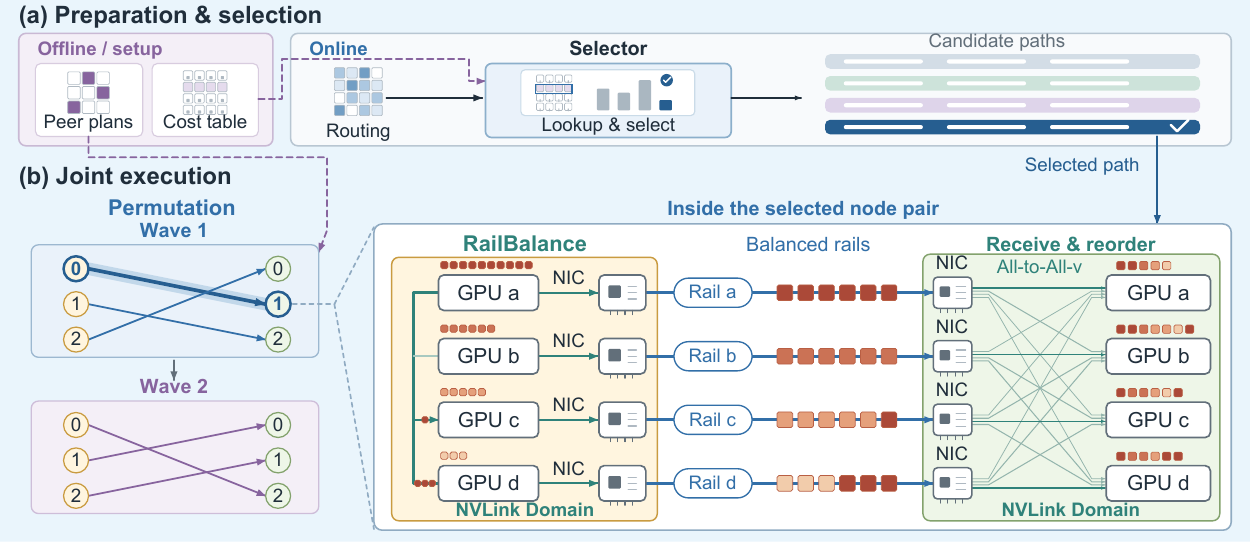}
    \caption{RailWave overview. Offline profiling and current-phase traffic
    statistics drive execution selection, while topology-derived peer plans are
    reused across phases. The highlighted Joint path applies RailBalance within
    scheduled node pairs.}
    \label{fig:RailWave-overview}
\end{figure}

RailWave combines source-local Rail redistribution with node-pair scheduling
on top of DeepEP. Figure~\ref{fig:RailWave-overview} shows its three-stage
workflow: preparation, selection, and execution. Offline profiling produces
a cost table, while initialization constructs reusable peer plans from the
EP topology. During execution, routing and window-derived traffic statistics
feed a selector that chooses among \textsc{Native}, \textsc{Rail-only},
\textsc{Permutation-only}, and \textsc{Joint}. Native retains the backend
execution path; Rail-only changes the egress allocation; Permutation-only
changes which node pairs communicate concurrently; Joint applies RailBalance
within each scheduled wave. The selector reuses validated offline decisions
near calibrated operating points and otherwise falls back to Joint.

The selected plan records the transfer order and owner-to-proxy mapping.
Dispatch delivers the original expert inputs through this plan, and combine
uses the saved metadata to return outputs to their original token positions.
All ranks follow the same active execution path and wave order, and the
selected plan remains fixed for the corresponding dispatch--combine pair.

\subsection{Source-Local RailBalance}
\label{sec:RailWave-rail}

RailBalance redistributes the traffic of each node pair $(s,d)$ across its
eligible Rails $\mathcal{R}_{s,d}$. It uses the source counts $C_{s,r,d}$
and fixed egress--ingress correspondence $\chi_{s,d}$ from
Section~\ref{sec:prelim-rails}. If source Rail $r$ carries $x_{s,r,d}$
transfer records, its contribution to the corresponding destination Rail is
{\small
\begin{equation}
\begin{aligned}
    I_{s,d,\chi_{s,d}(r)}
    &= x_{s,r,d}.
\end{aligned}
\label{eq:source-only-rail}
\end{equation}
}

The source's ingress contribution is therefore a permutation of its egress
allocation. Source-local balancing shapes both ends of a transfer without
a destination-side quota exchange. This guarantee concerns each source's
contribution; aggregate equality also depends on the eligible sets and
the distribution of integer remainders across sources.

Let $B_{s,d}=\sum_r C_{s,r,d}$ be the record count and
$m_{s,d}=|\mathcal{R}_{s,d}|$. RailBalance assigns quotas satisfying
{\small
\begin{equation}
\begin{aligned}
    Q_{s,r,d}
    &\in
    \left\{
        \left\lfloor\frac{B_{s,d}}{m_{s,d}}\right\rfloor,
        \left\lceil\frac{B_{s,d}}{m_{s,d}}\right\rceil
    \right\},
    \qquad
    \sum_{r\in\mathcal{R}_{s,d}} Q_{s,r,d}
    = B_{s,d}.
\end{aligned}
\label{eq:rail-quota}
\end{equation}
}

Quotas are zero outside the eligible set. Remainder ownership rotates
deterministically across destination slices, and eligible quotas differ
by at most one record. Counts retain the backend's per-destination
transfer accounting, including any existing token coalescing.

Records within quota remain on their owner Rail.
Algorithm~\ref{alg:railbalance} deterministically matches surplus records
to deficit slots using the shared count and quota vectors, yielding the
owner-to-proxy mapping without atomic slot allocation. Only nonidentity
mappings require relocation; token positions and expert metadata are retained
for combine.

\begin{algorithm}[t]
\caption{Source-local RailBalance plan construction}
\label{alg:railbalance}
\small
\begin{algorithmic}[1]
\Require Source counts $C_{s,r,d}$, eligible Rails $\mathcal R_{s,d}$
\Ensure Owner-to-proxy mapping $M_{s,d}$
\State $B_{s,d}\gets\sum_r C_{s,r,d}$
\State $m_{s,d}\gets|\mathcal R_{s,d}|$
\State $Q_{s,:,d}\gets\Call{BalancedQuotas}{B_{s,d},\mathcal R_{s,d},d}$
\Comment{Deterministic remainder rotation}
\State $\mathcal S\gets\emptyset$, $\mathcal D\gets\emptyset$
\For{each $r\in\mathcal R_{s,d}$}
    \State Keep the first $\min(C_{s,r,d},Q_{s,r,d})$ records on $r$
    \If{$C_{s,r,d}>Q_{s,r,d}$}
        \State Append excess records on $r$ to $\mathcal S$
    \ElsIf{$C_{s,r,d}<Q_{s,r,d}$}
        \State Append deficit slots on $r$ to $\mathcal D$
    \EndIf
\EndFor
\State Sort $\mathcal S$ by (owner Rail, local index)
\State Sort $\mathcal D$ by (target Rail, slot index)
\For{$i=1,\ldots,|\mathcal S|$}
    \State Map surplus record $\mathcal S[i]$ to the target Rail of $\mathcal D[i]$
\EndFor
\State \Return $M_{s,d}$
\end{algorithmic}
\end{algorithm}

\subsection{Reusable Permutation Scheduling}
\label{sec:RailWave-permutation}

Permutation scheduling bounds incast degree using cyclic pairwise
exchange~\citep{thakur2005optimization}. With $N$ nodes, wave
$k\in\{1,\ldots,N-1\}$ uses
{\small
\begin{equation}
\begin{aligned}
    \pi_k(s) &= (s+k)\bmod N, \\
    \pi_k^{-1}(d) &= (d-k)\bmod N.
\end{aligned}
\label{eq:cyclic-permutation}
\end{equation}
}

Each node has at most one outgoing and one incoming peer in a wave.
Every remote ordered pair appears exactly once because its shift is
uniquely determined by $(d-s)\bmod N$. These peer relationships depend
only on group membership and are constructed once; each phase supplies
new transfer records and counts.

For a wave, the executor selects records whose destination is
$\pi_k(s)$ and packs their payload without padding the node-pair demand.
This partition operates over backend transfer records, including multiple
destinations for the same token. Within each backend invocation, DeepEP
and Rail-only admit the submitted remote pairs together, whereas
Permutation-only and Joint execute the cached peer plans in order. Within each wave, disjoint node pairs communicate
concurrently.

Joint applies RailBalance to the records in the current wave. A phase
can be balanced across Rails in aggregate while a particular destination
slice remains concentrated on a few Rails. Redistributing that slice can
shorten the wave's critical transfer. The gain depends on its Rail skew
and relocation cost; scheduling still serializes the waves and retains
the original differences in node-pair demand.

\subsection{Calibrated Execution Selection}
\label{sec:RailWave-selector}

No single execution path dominates across communication phases.
RailBalance is beneficial when improved Rail utilization outweighs local
redistribution, while permutation scheduling is beneficial when reduced
incast outweighs the cost of additional communication rounds. These
break-even points vary with traffic characteristics, motivating
phase-specific execution selection.

The selector uses three traffic descriptors defined in
Section~\ref{sec:prelim-rails}: the average remote payload per GPU $B$,
the global source-Rail skew $\rho_g$, and the receiver concentration
$\kappa$ under the actual sending windows. Together, they capture
communication scale, source-Rail imbalance, and receiver concentration
for the current phase. After fixed normalization calibrated offline, these
statistics form the lookup feature
$x=(\widetilde B,\widetilde\rho_g,\widetilde\kappa)$.
Phases with no remote payload retain the Native local path.

Offline profiling constructs a calibrated anchor table. Each anchor stores
a feature vector $x_j$ and an execution path $p_j$. An alternative to Joint
is retained only when repeated measurements are sufficiently stable and
show a conservative advantage over Joint. This filters out decisions whose
apparent benefit may be explained by measurement variation. The exact
normalization, anchor-retention criteria, and resulting anchors are reported
in Appendix~\ref{app:adaptive-selector}.

At runtime, path decisions are reused only within a calibrated neighborhood
of these anchors. Let $\epsilon_m$ denote the neighborhood radius of feature
dimension $m$. The eligible anchor set is
{\small
\begin{equation}
\begin{aligned}
    \mathcal A(x)
    &=
    \left\{
        j:\,
        |x_m-x_{j,m}|\leq\epsilon_m,\;
        \forall m
    \right\}.
\end{aligned}
\label{eq:eligible-anchors}
\end{equation}
}

The runtime path decision is
{\small
\begin{equation}
\begin{aligned}
    p(x)
    &=
    \begin{cases}
        p_{j^*},
        & j^*=\displaystyle\arg\min_{j\in\mathcal A(x)}
          \|x-x_j\|_2^2, \\[2pt]
        \mathrm{Joint},
        & \mathcal A(x)=\varnothing.
    \end{cases}
\end{aligned}
\label{eq:calibrated-lookup}
\end{equation}
}

Thus, requests close to a validated operating point reuse its measured
path decision, while requests outside all calibrated neighborhoods
conservatively fall back to Joint. This guarded lookup avoids extrapolating
path choices into traffic regions not supported by offline measurements.

\section{Experiments}
\label{sec:evaluation}

We evaluate RailWave against Native, NCCL, and the released
FAST implementation~\citep{fast}, isolate the effects of spatial and
temporal shaping, and evaluate adaptive execution. Additional results
on integration with an upper-layer expert-parallel system are provided
in Appendix~\ref{app:ultraep}.

\subsection{Experimental Setup}
\label{sec:eval-setup}

\paragraph{Platforms.}
We evaluate on two 32-GPU clusters equipped with NVIDIA H800 and H20 GPUs,
respectively. Each cluster has four nodes with eight GPUs per node.
All compared methods use the same communication configuration within each
cluster. Detailed hardware and software settings are provided in
Appendix~\ref{app:system_config}.

\paragraph{Workloads.}
Our replay is based on expert-routing statistics collected during actual
GLM-4.5-Air~\citep{glm45} training runs on
DAPO-Math-17k~\citep{dapo_math_17k}, covering all 45 MoE layers.
To model incast under variable rank progress, we construct three controlled
sending patterns---Ring, Moderate, and Extreme---as detailed in
Appendix~\ref{app:replay-workloads}.

\paragraph{Baselines and metrics.}
We compare four execution paths on the same DeepEP backend:
\textsc{Native}, \textsc{Rail-only}, \textsc{Permutation-only}, and
\textsc{Joint}. Native and Rail-only retain the replayed transfer order,
whereas Permutation-only and Joint use cyclic waves; Joint additionally
applies RailBalance within each wave. RailWave selects among these paths using the calibrated lookup described
in Section~\ref{sec:RailWave-selector}. We also compare against NCCL
\texttt{all\_to\_all\_single} and the released FAST implementation.
The plotted latency is the GPU-event communication interval: all waves of
dispatch--combine for the DeepEP paths, or the corresponding forward--return
All-to-All for NCCL and FAST, including RailBalance GPU operations when
enabled. CPU planning, index construction, and prepacking are outside this
interval for every method, as detailed in
Appendix~\ref{app:replay-workloads}. Each plotted P50/P95/P99 is the sum of
the corresponding per-layer quantiles over the 45-layer replay.

\subsection{Overall Performance}
\label{sec:eval-overall}

\paragraph{Communication performance.}
Figures~\ref{fig:eval-overall-ring}--\ref{fig:eval-overall-extreme}
compare RailWave with Native, NCCL, and FAST under Ring, Moderate,
and Extreme, respectively. RailWave achieves $2.02$--$5.84\times$ speedups over Native
at P50 on H800 and $1.74$--$4.36\times$ on H20 across the three replay
patterns, with gains persisting at P95 and P99.

Native slows from Moderate to Extreme under stronger incast, whereas
RailWave maintains nearly identical P50 latency across the two patterns.

\begin{figure}[!htbp]
    \centering
    \captionsetup{font=small,skip=3pt,justification=raggedright,singlelinecheck=false}
\begin{minipage}[t]{0.49\textwidth}
        \centering\vspace{0pt}
        \RWEvalLegend{NCCL}{Native}{FAST}{RailWave}
        \RWEvalPairCap{evaluation_overall}{1}{2}{(a) H20}{(b) H800}
        \caption{Communication performance under Ring on H20 and H800.}
        \label{fig:eval-overall}
        \label{fig:eval-overall-ring}
    \end{minipage}\hfill%
    \begin{minipage}[t]{0.49\textwidth}
        \centering\vspace{0pt}
        \RWEvalLegend{NCCL}{Native}{FAST}{RailWave}
        \RWEvalPairCap{evaluation_overall}{3}{4}{(a) H20}{(b) H800}
        \caption{Communication performance under Moderate on H20 and H800.}
        \label{fig:eval-overall-moderate}
    \end{minipage}
\end{figure}

\subsection{Ablation and Mechanism Analysis}
\label{sec:eval-ablation}

\paragraph{Spatial and temporal contributions.}
Figure~\ref{fig:eval-components} separates the contributions of RailBalance
and permutation scheduling. In Ring, Rail-only closely matches Joint while
Permutation-only remains close to Native, showing that spatial shaping
provides most of the gain when incast is already limited.

Under Moderate and Extreme, combining the two mechanisms improves
performance on both clusters. On H800, Joint achieves $1.44\times$
and $1.97\times$ speedups over the better single mechanism,
respectively. RailBalance improves Rail utilization, whereas permutation
scheduling controls incast.

\begin{figure}[!htbp]
    \centering
    \captionsetup{font=small,skip=3pt,justification=raggedright,singlelinecheck=false}
\begin{minipage}[t]{0.49\textwidth}
        \centering\vspace{0pt}
        \RWEvalLegend{NCCL}{Native}{FAST}{RailWave}
        \RWEvalPairCap{evaluation_overall}{5}{6}{(a) H20}{(b) H800}
        \caption{Communication performance under Extreme on H20 and H800.}
        \label{fig:eval-overall-extreme}
    \end{minipage}\hfill%
    \begin{minipage}[t]{0.49\textwidth}
        \centering\vspace{0pt}
        \RWEvalLegend{Native}{Rail-only}{Permutation-only}{Joint}
        \RWEvalPairCap{evaluation_components}{1}{2}{(a) H20}{(b) H800}
        \caption{Ablation of RailBalance and permutation scheduling on H20 and H800.}
        \label{fig:eval-components}
    \end{minipage}
\end{figure}

\paragraph{Physical NIC load distribution.}
Figure~\ref{fig:eval-nic} shows the effect of RailBalance under Moderate
on H20. Native distributes transmitted traffic as approximately
54.12\%, 24.33\%, 14.85\%, and 6.69\% across the four NIC positions,
whereas Rail-only and Joint produce approximately equal shares in both
directions. Permutation-only retains the uneven distribution.

These results show that RailBalance changes physical Rail utilization,
whereas permutation scheduling changes which node-pair transfers overlap.

\paragraph{Performance across replay-derived traffic conditions.}
We further select cases
from the replay that span different levels of Rail skew and receiver
concentration. Figure~\ref{fig:eval-joint} shows that when both are high, Joint outperforms
either single mechanism. In one such case, Joint reduces mean latency to
4.11 ms, compared with 10.96 ms for Permutation-only and 73.28 ms for
Rail-only. More balanced cases can instead favor a lighter execution path,
motivating adaptive selection.

\begin{figure}[!htbp]
    \centering
    \captionsetup{font=small,skip=3pt,justification=raggedright,singlelinecheck=false}
\begin{minipage}[t]{0.49\textwidth}
        \centering\vspace{0pt}
        \RWEvalLegend{Native}{Rail-only}{Permutation-only}{Joint}
        \RWEvalPairCap{evaluation_nic}{1}{2}{(a) Transmit}{(b) Receive}
        \caption{Physical NIC traffic distribution under Moderate on H20.}
        \label{fig:eval-nic}
    \end{minipage}\hfill%
    \begin{minipage}[t]{0.49\textwidth}
        \centering\vspace{0pt}
        \RWEvalLegend{Native}{Rail-only}{Permutation-only}{Joint}
        \RWEvalPairCap{evaluation_joint}{1}{2}{(a) 14 MiB}{(b) 56 MiB}
        \caption{Execution-path performance across replay-derived cases with
        different Rail-skew and receiver-concentration characteristics.}
        \label{fig:eval-joint}
    \end{minipage}
\end{figure}

\begin{figure}[t]
    \centering
    \includegraphics[width=0.62\linewidth]{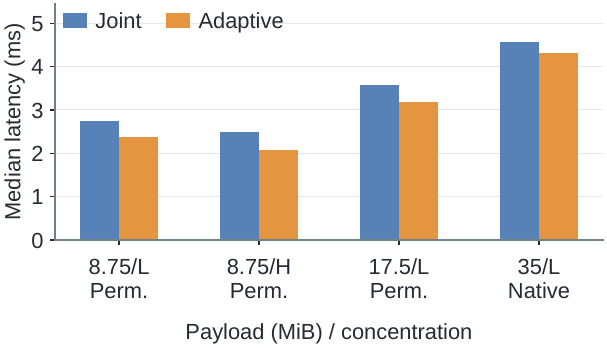}
    \caption{Adaptive execution on unseen requests where the calibrated
    selector switches away from Joint. L/H denote low/high receiver
    concentration; the selected execution path is shown below each case.}
    \label{fig:eval-adaptive}
\end{figure}

\paragraph{Adaptive execution.}
We evaluate the calibrated selector on 12 unseen requests that are not used
to construct the frozen cost table. Figure~\ref{fig:eval-adaptive} shows the four
requests where the selector switches away from Joint: three select
Permutation-only and one selects Native. All four switches reduce measured
median communication latency, by 5.7--16.4\%, corresponding to
$1.061$--$1.196\times$ speedup over Joint. The remaining eight requests
fall back to Joint. Complete calibration and validation results, including
the development-regression set, are provided in
Appendix~\ref{app:adaptive-selector}.

Additional UltraEP integration results, packing measurements, and workspace
accounting are provided in Appendices~\ref{app:ultraep},
\ref{app:packing_overhead}, and~\ref{app:memory_overhead}, respectively.

\section{Conclusion} We presented RailWave, a communication layer for expert-parallel MoE execution that targets physical-network inefficiencies remaining after expert routing and placement are fixed. RailWave combines source-local RailBalance with reusable permutation scheduling to jointly improve parallel Rail utilization and control receiver-side concurrency, while preserving expert assignments and logical communication demand. Built on DeepEP, RailWave achieves communication speedups of up to $5.84\times$ on H800 and $4.36\times$ on H20, demonstrating the benefits of jointly shaping the spatial and temporal execution of expert-parallel traffic.

%

\subsection*{AI use statement}

In this work, we used generative AI tools to assist with language polishing,
code debugging, and figure revision. All AI-assisted text was reviewed and
edited by the authors, code changes were inspected and tested, and figure
revisions were verified against the underlying data. We take responsibility
for the final content of this work, including all text, code, figures, claims,
and other artifacts produced with the aid of generative AI.

\subsection*{Reproducibility statement}

We provide the design and implementation details of RailWave in
Section~\ref{sec:system}, and describe the evaluation setup and workloads in
Section~\ref{sec:eval-setup}. Further details cover hardware and software
settings (Appendix~\ref{app:system_config}), UltraEP integration
(Appendix~\ref{app:ultraep}), selector calibration and validation
(Appendix~\ref{app:adaptive-selector}), packing overhead
(Appendix~\ref{app:packing_overhead}), memory overhead
(Appendix~\ref{app:memory_overhead}), and the replay protocol
(Appendix~\ref{app:replay-workloads}). An anonymized
source-code repository will be provided with the submission to support
reproduction of the reported results. The repository will be made public
upon publication.

\bibliography{iclr2027_conference}

@techreport{nvidia2023superpod,
  author      = {{NVIDIA}},
  title       = {{NVIDIA DGX SuperPOD}: Next Generation Scalable Infrastructure for {AI} Leadership: Reference Architecture Featuring {NVIDIA DGX H100} Systems},
  institution = {NVIDIA},
  number      = {RA-11333-001 V11},
  year        = {2023},
  month       = sep,
  url         = {https://docs.nvidia.com/dgx-superpod/reference-architecture-scalable-infrastructure-h100/latest/}
}

@article{deepseekv3,
  title   = {{DeepSeek-V3} Technical Report},
  author  = {{DeepSeek-AI}},
  journal = {arXiv preprint arXiv:2412.19437},
  year    = {2024},
  url     = {https://arxiv.org/abs/2412.19437}
}

@article{glm45,
  title   = {{GLM-4.5}: Agentic, Reasoning, and Coding ({ARC}) Foundation Models},
  author  = {{GLM-4.5 Team}},
  journal = {arXiv preprint arXiv:2508.06471},
  year    = {2025},
  url     = {https://arxiv.org/abs/2508.06471}
}

@article{dapo_math_17k,
  title   = {{DAPO}: An Open-Source {LLM} Reinforcement Learning System at Scale},
  author  = {Yu, Qiying and Zhang, Zheng and Zhu, Ruofei and Yuan, Yufeng and
             Zuo, Xiaochen and Yue, Yu and Dai, Weinan and Fan, Tiantian and Liu, Gaohong and
             Liu, Lingjun and Liu, Xin and Lin, Haibin and Lin, Zhiqi and
             Ma, Bole and Sheng, Guangming and Tong, Yuxuan and Zhang, Chi and
             Zhang, Mofan and Zhang, Wang and Zhu, Hang and Zhu, Jinhua and
             Chen, Jiaze and Chen, Jiangjie and Wang, Chengyi and Yu, Hongli and
             Song, Yuxuan and Wei, Xiangpeng and Zhou, Hao and
             Liu, Jingjing and Ma, Wei-Ying and Zhang, Ya-Qin and Yan, Lin and
             Qiao, Mu and Wu, Yonghui and Wang, Mingxuan},
  journal = {arXiv preprint arXiv:2503.14476},
  year    = {2025},
  url     = {https://arxiv.org/abs/2503.14476}
}

@inproceedings{shazeer2017outrageously,
  title     = {Outrageously Large Neural Networks: The Sparsely-Gated Mixture-of-Experts Layer},
  author    = {Shazeer, Noam and Mirhoseini, Azalia and Maziarz, Krzysztof and
               Davis, Andy and Le, Quoc and Hinton, Geoffrey and Dean, Jeff},
  booktitle = {International Conference on Learning Representations},
  year      = {2017},
  url       = {https://openreview.net/forum?id=B1ckMDqlg}
}

@inproceedings{lepikhin2021gshard,
  title     = {{GShard}: Scaling Giant Models with Conditional Computation and Automatic Sharding},
  author    = {Lepikhin, Dmitry and Lee, HyoukJoong and Xu, Yuanzhong and
               Chen, Dehao and Firat, Orhan and Huang, Yanping and Krikun, Maxim and
               Shazeer, Noam and Chen, Zhifeng},
  booktitle = {International Conference on Learning Representations},
  year      = {2021},
  url       = {https://arxiv.org/abs/2006.16668}
}

@article{fedus2022switch,
  title   = {Switch Transformers: Scaling to Trillion Parameter Models with Simple and Efficient Sparsity},
  author  = {Fedus, William and Zoph, Barret and Shazeer, Noam},
  journal = {Journal of Machine Learning Research},
  volume  = {23},
  number  = {120},
  pages   = {1--39},
  year    = {2022},
  url     = {https://www.jmlr.org/papers/v23/21-0998.html}
}

@inproceedings{liao2025mixnet,
  title     = {{MixNet}: A Runtime Reconfigurable Optical-Electrical Fabric for Distributed Mixture-of-Experts Training},
  author    = {Liao, Xudong and Sun, Yijun and Tian, Han and Wan, Xinchen and
               Jin, Yilun and Wang, Zilong and Ren, Zhenghang and Huang, Xinyang and
               Li, Wenxue and Tse, Kin Fai and Zhong, Zhizhen and Liu, Guyue and
               Zhang, Ying and Ye, Xiaofeng and Zhang, Yiming and Chen, Kai},
  booktitle = {Proceedings of the ACM SIGCOMM 2025 Conference},
  pages     = {554--574},
  year      = {2025},
  doi       = {10.1145/3718958.3750465},
  url       = {https://doi.org/10.1145/3718958.3750465}
}

@inproceedings{li2023accelerating,
  title     = {Accelerating Distributed {MoE} Training and Inference with {Lina}},
  author    = {Li, Jiamin and Jiang, Yimin and Zhu, Yibo and Wang, Cong and Xu, Hong},
  booktitle = {2023 USENIX Annual Technical Conference (USENIX ATC 23)},
  pages     = {945--959},
  year      = {2023},
  month     = jul,
  publisher = {USENIX Association},
  address   = {Boston, MA},
  isbn      = {978-1-939133-35-9},
  url       = {https://www.usenix.org/conference/atc23/presentation/li-jiamin}
}

@inproceedings{lewis2021base,
  title     = {{BASE} Layers: Simplifying Training of Large, Sparse Models},
  author    = {Lewis, Mike and Bhosale, Shruti and Dettmers, Tim and
               Goyal, Naman and Zettlemoyer, Luke},
  booktitle = {Proceedings of the 38th International Conference on Machine Learning},
  series    = {Proceedings of Machine Learning Research},
  volume    = {139},
  pages     = {6265--6274},
  year      = {2021},
  publisher = {PMLR},
  url       = {https://proceedings.mlr.press/v139/lewis21a.html}
}

@inproceedings{zhou2022mixture,
  title     = {Mixture-of-Experts with Expert Choice Routing},
  author    = {Zhou, Yanqi and Lei, Tao and Liu, Hanxiao and Du, Nan and
               Huang, Yanping and Zhao, Vincent and Dai, Andrew M. and
               Chen, Zhifeng and Le, Quoc V. and Laudon, James},
  booktitle = {Advances in Neural Information Processing Systems},
  volume    = {35},
  pages     = {7103--7114},
  year      = {2022},
  url       = {https://arxiv.org/abs/2202.09368}
}

@article{nie2023flexmoe,
  title   = {{FlexMoE}: Scaling Large-Scale Sparse Pre-Trained Model Training via Dynamic Device Placement},
  author  = {Nie, Xiaonan and Miao, Xupeng and Wang, Zilong and Yang, Zichao and
             Xue, Jilong and Ma, Lingxiao and Cao, Gang and Cui, Bin},
  journal = {Proceedings of the ACM on Management of Data},
  volume  = {1},
  number  = {1},
  pages   = {1--19},
  year    = {2023},
  doi     = {10.1145/3588964},
  url     = {https://doi.org/10.1145/3588964}
}

@inproceedings{he2022fastermoe,
  title     = {{FasterMoE}: Modeling and Optimizing Training of Large-Scale Dynamic Pre-Trained Models},
  author    = {He, Jiaao and Zhai, Jidong and Antunes, Tiago and Wang, Haojie and
               Luo, Fuwen and Shi, Shangfeng and Li, Qin},
  booktitle = {Proceedings of the 27th ACM SIGPLAN Symposium on Principles and Practice of Parallel Programming},
  pages     = {120--134},
  year      = {2022},
  doi       = {10.1145/3503221.3508418},
  url       = {https://doi.org/10.1145/3503221.3508418}
}

@misc{eplb,
  title        = {{EPLB}: Expert Parallelism Load Balancer},
  author       = {{DeepSeek-AI}},
  year         = {2025},
  howpublished = {\url{https://github.com/deepseek-ai/EPLB}}
}

@article{wei2026ultraep,
  title   = {{UltraEP}: Unleash {MoE} Training and Inference on Rack-Scale Nodes with Near-Optimal Load Balancing},
  author  = {Wei, Xinming and Jin, Chao and Dai, Tuo and Zhong, Yinmin and
             Yu, Shan and Yang, Chengxu and Wu, Bingyang and Zhang, Zili and
             Mai, Jing and Zhu, Qianchao and Li, Zhouyang and Liu, Yuliang and
             Luo, Guojie},
  journal = {arXiv preprint arXiv:2606.04101},
  year    = {2026},
  url     = {https://arxiv.org/abs/2606.04101}
}

@misc{moonshot2026moonep,
  title        = {{MoonEP}: A Perfectly Balanced Expert Parallelism Library via Dynamic Redundant Experts},
  author       = {Chen, Yutian and Li, Cong and Wang, Yucheng and Wei, Ming},
  year         = {2026},
  howpublished = {\url{https://github.com/MoonshotAI/MoonEP}}
}

@article{zhang2026taot,
  title   = {{TAOT}: Topology-Aware Optimal Transport for Dynamic Expert Replica Placement in {MoE} Training},
  author  = {Zhang, Lingyun and Zhang, Henghua and Gu, Shilei and Mo, Kai and
             Han, Shuai and Li, Shiyong and Wang, Yanpeng and Shen, Dou},
  journal = {arXiv preprint arXiv:2608.03676},
  year    = {2026},
  url     = {https://arxiv.org/abs/2608.03676}
}

@inproceedings{liu2025netmoe,
  title     = {{NetMoE}: Accelerating {MoE} Training through Dynamic Sample Placement},
  author    = {Liu, Xinyi and Wang, Yujie and Fu, Fangcheng and Miao, Xupeng and
               Zhu, Shenhan and Nie, Xiaonan and Cui, Bin},
  booktitle = {International Conference on Learning Representations},
  year      = {2025},
  url       = {https://proceedings.iclr.cc/paper_files/paper/2025/hash/e0c256700465c158de71081b4cf5e8c3-Abstract-Conference.html}
}

@inproceedings{wang2024railonly,
  title     = {Rail-only: A Low-Cost High-Performance Network for Training {LLMs} with Trillion Parameters},
  author    = {Wang, Weiyang and Ghobadi, Manya and Shakeri, Kayvon and
               Zhang, Ying and Hasani, Naader},
  booktitle = {2024 IEEE Symposium on High-Performance Interconnects (HOTI)},
  pages     = {1--10},
  year      = {2024},
  publisher = {IEEE},
  doi       = {10.1109/HOTI63208.2024.00013},
  url       = {https://doi.org/10.1109/HOTI63208.2024.00013}
}

@article{rails,
  title   = {{RailS}: Load Balancing for All-to-All Communication in Distributed Mixture-of-Experts Training},
  author  = {Xu, Heng and Yu, Zhiwei and Du, Chengze and Zhou, Ying and
             Li, Letian and Wang, Haojie and Cheng, Weiqiang and Li, Jialong},
  journal = {IEEE Transactions on Networking},
  volume  = {34},
  pages   = {4431--4448},
  year    = {2026},
  doi     = {10.1109/TON.2026.3676382},
  url     = {https://doi.org/10.1109/TON.2026.3676382}
}

@inproceedings{fuselink,
  title     = {Enabling Efficient {GPU} Communication over Multiple {NICs} with {FuseLink}},
  author    = {Ren, Zhenghang and Li, Yuxuan and Wang, Zilong and Huang, Xinyang and
               Li, Wenxue and Xu, Kaiqiang and Liao, Xudong and Sun, Yijun and
               Liu, Bowen and Tian, Han and Zhang, Junxue and Wang, Mingfei and
               Zhong, Zhizhen and Liu, Guyue and Zhang, Ying and Chen, Kai},
  booktitle = {19th USENIX Symposium on Operating Systems Design and Implementation (OSDI 25)},
  pages     = {91--108},
  year      = {2025},
  month     = jul,
  publisher = {USENIX Association},
  address   = {Boston, MA},
  isbn      = {978-1-939133-47-2},
  url       = {https://www.usenix.org/conference/osdi25/presentation/ren}
}

@inproceedings{sabre,
  title     = {Skew-aware Adaptive All-to-allv Algorithms for Dynamic Deep Learning Workloads},
  author    = {Wei, Cunyang and Bhatele, Abhinav},
  booktitle = {Proceedings of the 40th ACM International Conference on Supercomputing},
  pages     = {450--462},
  year      = {2026},
  doi       = {10.1145/3797905.3800541},
  url       = {https://doi.org/10.1145/3797905.3800541}
}

@inproceedings{fast,
  title     = {{FAST}: An Efficient Scheduler for {All-to-All} {GPU} Communication},
  author    = {Lei, Yiran and Lee, Dongjoo and Zhao, Liangyu and Kurniawan, Daniar and
               Kim, Chanmyeong and Jeong, Heetaek and Kim, Changsu and Choi, Hyeonseong and
               Yu, Liangcheng and Krishnamurthy, Arvind and Sherry, Justine and
               Nurvitadhi, Eriko},
  booktitle = {23rd USENIX Symposium on Networked Systems Design and Implementation (NSDI 26)},
  pages     = {2515--2531},
  year      = {2026},
  month     = may,
  publisher = {USENIX Association},
  address   = {Renton, WA},
  isbn      = {978-1-939133-54-0},
  url       = {https://www.usenix.org/conference/nsdi26/presentation/lei-yiran}
}

@article{wu2026hierarchicalbvn,
  title   = {Dynamic Hierarchical {Birkhoff--von Neumann} Decomposition for {All-to-All} {GPU} Communication},
  author  = {Wu, Yen-Chieh and Chang, Cheng-Shang and Lee, Duan-Shin and Chao, H. Jonathan},
  journal = {arXiv preprint arXiv:2602.22756},
  year    = {2026},
  url     = {https://arxiv.org/abs/2602.22756}
}

@article{incastfree,
  title   = {Incast-Free {MoE} Rate-Based Scheduling},
  author  = {Cohen, Evyatar and Yallouz, Jose and Shpiner, Alexander and
             Silberstein, Mark and Ratnasamy, Sylvia and Keslassy, Isaac},
  journal = {arXiv preprint arXiv:2607.26340},
  year    = {2026},
  url     = {https://arxiv.org/abs/2607.26340}
}

@article{thakur2005optimization,
  title   = {Optimization of Collective Communication Operations in {MPICH}},
  author  = {Thakur, Rajeev and Rabenseifner, Rolf and Gropp, William},
  journal = {The International Journal of High Performance Computing Applications},
  volume  = {19},
  number  = {1},
  pages   = {49--66},
  year    = {2005},
  doi     = {10.1177/1094342005051521},
  url     = {https://doi.org/10.1177/1094342005051521}
}

@misc{deepseek2026deepep,
  title        = {{DeepEP}: An Efficient Expert-Parallel Communication Library},
  author       = {Zhao, Chenggang and Zhou, Shangyan and Zhang, Liyue and Deng, Chengqi and
                  Xu, Zhean and Liu, Yuxuan and Yu, Kuai and Li, Jiashi and Zhao, Liang},
  year         = {2025},
  howpublished = {\url{https://github.com/deepseek-ai/DeepEP}},
  note         = {Official upstream package, version 2.1.0}
}

@article{goldman2026ncclep,
  title   = {{NCCL EP}: Towards a Unified Expert Parallel Communication {API} for {NCCL}},
  author  = {Goldman, Amos and Boker, Nimrod and Sheraizin, Maayan and Admoni, Nimrod and
             Polyakov, Artem and Bhattacharya, Subhadeep and Yu, Fan and Sun, Kai and
             Theodorakis, Georgios and Yin, Hsin-Chun and Gootzen, Peter-Jan and
             Shafi, Aamir and Ravid, Assaf and Di Girolamo, Salvatore and
             Dinan, James and Li, Xiaofan and Venkata, Manjunath Gorentla and
             Bloch, Gil},
  journal = {arXiv preprint arXiv:2603.13606},
  year    = {2026},
  url     = {https://arxiv.org/abs/2603.13606}
}

@inproceedings{li2026swiftep,
  title     = {{SwiftEP}: Accelerating {MoE} Inference with Buffer Fusion and {TMA} Offloading},
  author    = {Li, Xingyi and Liu, Yadong and Huang, Xiaojie and Zhang, Yiran and
               Wang, Shuai and Wang, Shangguang and Lin, Zhehao and Xia, Yinben and
               Yu, Chang and Liu, Qihang and Zhang, Xuan and Lu, Hao and Li, Xiang and
               He, Zekun and Wang, Yachen and Zou, Xianneng},
  booktitle = {23rd USENIX Symposium on Networked Systems Design and Implementation (NSDI 26)},
  pages     = {1073--1089},
  year      = {2026},
  month     = may,
  publisher = {USENIX Association},
  address   = {Renton, WA},
  isbn      = {978-1-939133-54-0},
  url       = {https://www.usenix.org/conference/nsdi26/presentation/li-xingyi}
}

@inproceedings{liu2026ubep,
  title     = {{UBEP}: Re-architecting Expert Parallelism Communication Library for Production Superpods},
  author    = {Liu, Yipeng and Liu, Chang and Shen, Si and Zheng, Jiaqi and
               Li, Mingfan and Yang, Yuyang and Li, Guanhua and Zhang, Yuquan and
               Xu, Yimeng and Hu, Zhongzhe and Huang, Zhiyuan and Duan, Qihang and
               Wang, Junsong and Ling, Wenkai and Yang, Baochuan and Yu, Xianzhi and
               Bao, Han and Chen, Yijie and Chen, Guihai},
  booktitle = {Proceedings of the ACM SIGCOMM 2026 Conference},
  pages     = {415--432},
  year      = {2026},
  doi       = {10.1145/3789240.3829183},
  url       = {https://doi.org/10.1145/3789240.3829183}
}

@inproceedings{cao2026balancing,
  title     = {Balancing and Beyond: Communication-Centric Optimizations in Expert Parallelism},
  author    = {Cao, Jiamin and Li, Qingxu and Liu, Yaozhong and Gao, Jiaqi and
               Zhang, Yan and Shi, Shangfeng and Chen, Zian and Wang, Yizhi and
               Zhang, Jun and He, Kunling and Zhai, Ennan and Dong, Jianbo and
               Fu, Binzhang and Cai, Dennis},
  booktitle = {Proceedings of the ACM SIGCOMM 2026 Conference},
  pages     = {550--563},
  year      = {2026},
  doi       = {10.1145/3789240.3829201},
  url       = {https://doi.org/10.1145/3789240.3829201}
}

@article{nie2024lsh,
  title   = {{LSH-MoE}: Communication-Efficient {MoE} Training via Locality-Sensitive Hashing},
  author  = {Nie, Xiaonan and Liu, Qibin and Fu, Fangcheng and Zhu, Shenhan and
             Miao, Xupeng and Li, Xiaoyang and Zhang, Yang and Liu, Shouda and Cui, Bin},
  journal = {Advances in Neural Information Processing Systems},
  volume  = {37},
  pages   = {54161--54182},
  year    = {2024},
  doi     = {10.52202/079017-1716},
  url     = {https://proceedings.neurips.cc/paper_files/paper/2024/hash/61674667d642ae52f6bb281bea90ee29-Abstract-Conference.html}
}

@article{jiang2024lancet,
  title   = {{Lancet}: Accelerating Mixture-of-Experts Training via Whole Graph Computation-Communication Overlapping},
  author  = {Jiang, Chenyu and Tian, Ye and Jia, Zhen and Zheng, Shuai and Wu, Chuan and Wang, Yida},
  journal = {Proceedings of Machine Learning and Systems},
  volume  = {6},
  pages   = {74--86},
  year    = {2024},
  url     = {https://proceedings.mlsys.org/paper_files/paper/2024/hash/339caf45a6fa281cae8adc6465343464-Abstract-Conference.html}
}

@article{zhang2025comet,
  title   = {{COMET}: Fine-Grained Computation-Communication Overlapping for Mixture-of-Experts},
  author  = {Zhang, Shulai and Zheng, Ningxin and Lin, Haibin and Jiang, Ziheng and
             Bao, Wenlei and Jiang, Chengquan and Hou, Qi and Cui, Weihao and
             Zheng, Size and Chang, Li-Wen and Chen, Quan and Liu, Xin},
  journal = {Proceedings of Machine Learning and Systems},
  volume  = {7},
  year    = {2025},
  url     = {https://proceedings.mlsys.org/paper_files/paper/2025/hash/e27ea0cd50b798ff8942caf9203f0992-Abstract-Conference.html}
}

@article{shoeybi2020megatron,
  author  = {Mohammad Shoeybi and Mostofa Patwary and Raul Puri and Patrick LeGresley and Jared Casper and Bryan Catanzaro},
  title   = {{Megatron-LM}: Training Multi-Billion Parameter Language Models Using Model Parallelism},
  journal = {arXiv preprint arXiv:1909.08053},
  year    = {2020},
  note    = {Version 4},
  url     = {https://arxiv.org/abs/1909.08053v4}
}

@article{xhebraj2024jaxpp,
  author  = {Anxhelo Xhebraj and Sean Lee and Hanfeng Chen and Vinod Grover},
  title   = {Scaling Deep Learning Training with {MPMD} Pipeline Parallelism},
  journal = {arXiv preprint arXiv:2412.14374},
  year    = {2024},
  note    = {Version 1},
  url     = {https://arxiv.org/abs/2412.14374v1}
}
\bibliographystyle{iclr2027_conference}

\newpage

\appendix
\section{Implementations}
\label{app:appendix}
\subsection{Hardware and Software Configuration}
\label{app:system_config}

Table~\ref{tab:system_config} summarizes the hardware and software
configurations of the two clusters used in our evaluation.
Both clusters contain four nodes with eight GPUs per node.
Unless otherwise specified, the software versions correspond to the
real-routing replay experiments.

\begin{table}[h]
\centering
\caption{Hardware and software configurations of the evaluation clusters.}
\label{tab:system_config}
\resizebox{\linewidth}{!}{
\begin{tabular}{lll}
\toprule
Configuration &
H800 Cluster &
H20 Cluster \\
\midrule

Number of nodes &
4 &
4 \\

GPUs per node &
8 &
8 \\

GPU model &
NVIDIA H800 &
NVIDIA H20 \\

GPU memory &
81,559 MiB &
97,871 MiB \\

Intra-node interconnect &
NVLink &
NVLink\\

Number of NICs per node &
8 RDMA HCA  &
4 RDMA HCA \\

NIC bandwidth &
200 Gb/s per port &
200 Gb/s per port \\

Inter-node network &
RDMA; physical fabric type not recorded &
RDMA; physical fabric type not recorded \\

\midrule

NVIDIA driver &
550.90.07 &
590.48.01 \\

CUDA &
12.9  &
12.8  \\

NCCL &
2.30.7 &
2.31.2 \\

DeepEP &
Official upstream base, package v2.1.0 &
Official upstream base, package v2.1.0 \\

\midrule

GPU--NIC mapping &
One PIX-proximate HCA per GPU &
NCCL-managed\\

RDMA backend &
NCCL GIN, RAIL connection &
NCCL GIN, RAIL connection \\

DeepEP communication contexts &
129 GIN contexts &
129 GIN contexts \\

Communication SMs &
16 &
16 \\
\bottomrule
\end{tabular}}
\end{table}

Each H800 node has eight 400-Gb/s NVIDIA ConnectX-7 adapters, with one PIX-proximate adapter per GPU. Each H20 node has four 400-Gb/s NVIDIA ConnectX-7 adapters, each configured with two 200-Gb/s ports. Each pair of PCIe-local GPUs shares one adapter, with a separate port assigned to each GPU for port-level traffic isolation.

Native, Rail-only, Permutation-only, and Joint use the same backend,
based on the official DeepEP package version 2.1.0, and the same launch-level
communication configuration within each
paired replay experiment. The DeepEP configuration allocates 129 GIN
communication contexts and uses 16 communication SMs with automatic
QP selection. External baselines such as NCCL and FAST retain their
own communication backends and library-specific settings.

\subsection{Integration with UltraEP}
\label{app:ultraep}

We further evaluate RailWave when integrated with UltraEP. UltraEP determines expert placement and maps logical expert IDs to physical experts, while RailWave operates below the routing layer and optimizes how the resulting token traffic is executed on the network. RailWave does not modify the physical expert selected by UltraEP.

For each MoE layer, we first invoke the UltraEP manager to perform expert placement, weight synchronization, and logical-to-physical expert rerouting. The resulting physical expert IDs and routing weights are then passed to the DeepEP communication backend. All compared communication modes use the same UltraEP routing results.

\paragraph{Workloads.}
We construct two synthetic traffic families, both with three source nodes sending to the same destination node. The first introduces \emph{source-node skew}: traffic volumes differ across the three source nodes, while traffic within each source node is approximately balanced across its local Rails. This workload primarily exposes receiver-side contention.

The second introduces \emph{source-Rail skew}: the three source nodes contribute approximately equal traffic volumes, while traffic within each source node is concentrated on a subset of Rails. This workload therefore combines receiver-side contention with source-Rail imbalance.

We evaluate two traffic volumes, corresponding to 65,536 and 98,304 remote tokens. With BF16 activations and a hidden dimension of 7,168, they correspond to average inter-node payloads of approximately 28 MiB and 42 MiB per GPU per All-to-All operation, respectively. Logical routing is approximately uniform across the eight experts on the destination node.

UltraEP placement and rerouting are performed before communication. In these workloads, the hotspot experts are already distributed within the destination node, so UltraEP does not move the resulting traffic away from that node.

\paragraph{Timing boundary.}
The reported communication latency includes DeepEP dispatch and combine, as well as the RailBalance GPU operations when enabled. UltraEP placement, weight synchronization, and rerouting are measured separately and are not included in the communication latency. Destination grouping, prepacking, and output-order reconstruction are also performed outside the communication timing interval.

\paragraph{Communication results.}
Table~\ref{tab:ultraep_comm} reports the communication latency under the two traffic families.

For source-node-skew traffic, temporal scheduling provides most of the benefit. At 28 MiB per GPU, Permutation reduces p50 latency from 108.478 ms to 11.191 ms, corresponding to a $9.694\times$ speedup over Native. At 42 MiB, it reduces latency from 141.361 ms to 16.305 ms, corresponding to an $8.670\times$ speedup. Since the source-side Rails are already approximately balanced in this workload, applying RailBalance on top of permutation scheduling provides little additional benefit.

For source-Rail-skew traffic, spatial and temporal traffic shaping become complementary. At 28 MiB, Joint reduces p50 latency from 138.624 ms to 17.632 ms, corresponding to a $7.862\times$ speedup over Native. At 42 MiB, Joint reduces latency from 233.130 ms to 25.721 ms, corresponding to a $9.064\times$ speedup. Compared with the better of RailOnly and Permutation, Joint provides additional speedups of $2.125\times$ and $3.295\times$ at 28 MiB and 42 MiB, respectively.

These results show that the preferred execution strategy depends on the traffic pattern. Temporal scheduling is sufficient when the Rails are already balanced, whereas jointly applying spatial and temporal shaping provides substantially larger gains when Rail imbalance and receiver contention coexist.

\begin{table}[t]
\centering
\small
\caption{UltraEP integration results under source-node skew (N) and source-Rail skew (R). 
28 and 42 denote the average inter-node payload in MiB per GPU per All-to-All operation.
Speedup is relative to Native using p50 latency. Best values in each case are bold.}
\label{tab:ultraep_comm}
\setlength{\tabcolsep}{4pt}
\begin{tabular}{llrrr}
\toprule
Case & Mode & p50 (ms) & p99 (ms) & Speedup \\
\midrule
N28 & Native & 108.478 & 154.055 & 1.000$\times$ \\
N28 & RailOnly & 79.735 & 153.605 & 1.360$\times$ \\
N28 & Permutation & {\bfseries\boldmath 11.191} & 12.440 & {\bfseries\boldmath 9.694$\times$} \\
N28 & Joint & 11.489 & {\bfseries\boldmath 12.138} & 9.442$\times$ \\
\midrule
N42 & Native & 141.361 & 192.259 & 1.000$\times$ \\
N42 & RailOnly & 76.374 & 93.600 & 1.851$\times$ \\
N42 & Permutation & {\bfseries\boldmath 16.305} & 19.180 & {\bfseries\boldmath 8.670$\times$} \\
N42 & Joint & 16.712 & {\bfseries\boldmath 17.942} & 8.459$\times$ \\
\midrule
R28 & Native & 138.624 & 243.630 & 1.000$\times$ \\
R28 & RailOnly & 37.467 & 106.081 & 3.700$\times$ \\
R28 & Permutation & 56.899 & 59.076 & 2.436$\times$ \\
R28 & Joint & {\bfseries\boldmath 17.632} & {\bfseries\boldmath 19.506} & {\bfseries\boldmath 7.862$\times$} \\
\midrule
R42 & Native & 233.130 & 352.334 & 1.000$\times$ \\
R42 & RailOnly & 102.411 & 130.325 & 2.276$\times$ \\
R42 & Permutation & 84.742 & 84.892 & 2.751$\times$ \\
R42 & Joint & {\bfseries\boldmath 25.721} & {\bfseries\boldmath 25.849} & {\bfseries\boldmath 9.064$\times$} \\
\bottomrule
\end{tabular}
\end{table}

\paragraph{UltraEP control overhead.}
We separately measure the UltraEP placement, rerouting, and weight-synchronization overheads, as shown in Table~\ref{tab:ultraep_control}. Placement and rerouting remain on the order of tens of microseconds across the evaluated workloads and are small relative to the communication latency.

The weight-synchronization measurements are included only to verify the execution of the UltraEP integration path. The evaluated FC weight tensors contain only 1,024 BF16 elements and therefore do not represent the synchronization cost of full-size MoE expert weights.

\begin{table}[t]
\centering
\small
\caption{UltraEP control-operation latency in microseconds. These costs are measured separately and are not included in Table~\ref{tab:ultraep_comm}.}
\label{tab:ultraep_control}
\begin{tabular}{llrrr}
\toprule
Case & Mode & Placement & Reroute & Weight Sync \\
\midrule
N28 & Native      &  84.20 & 38.51 & 38.40 \\
N28 & RailOnly    &  87.17 & 55.02 & 44.59 \\
N28 & Permutation &  86.13 & 36.79 & 36.03 \\
N28 & Joint       &  84.97 & 37.84 & 37.23 \\
\midrule
N42 & Native      &  84.00 & 43.76 & 38.31 \\
N42 & RailOnly    &  85.72 & 39.70 & 47.92 \\
N42 & Permutation &  84.97 & 39.59 & 36.45 \\
N42 & Joint       &  85.46 & 52.76 & 37.23 \\
\midrule
R28 & Native      &  82.28 & 37.58 & 43.93 \\
R28 & RailOnly    &  84.78 & 39.94 & 36.87 \\
R28 & Permutation &  85.36 & 38.38 & 36.87 \\
R28 & Joint       &  85.83 & 38.82 & 48.34 \\
\midrule
R42 & Native      &  84.46 & 42.57 & 38.75 \\
R42 & RailOnly    &  85.53 & 41.55 & 42.10 \\
R42 & Permutation & 103.58 & 63.42 & 59.94 \\
R42 & Joint       &  87.91 & 45.16 & 54.48 \\
\bottomrule
\end{tabular}
\end{table}

\subsection{Adaptive Selector Calibration and Validation}
\label{app:adaptive-selector}

\paragraph{Calibration details.}
The selector uses
$x=(\widetilde B,\widetilde\rho_g,\widetilde\kappa)$, with
\[
\widetilde B=\log_2(B/56),\qquad
\widetilde\rho_g=(\rho_g-1)/7,\qquad
\widetilde\kappa=(\kappa-1)/2.
\]
These constants are fixed engineering normalization scales rather than
fitted model parameters. The 56-MiB reference corresponds to 32,768
tokens per node under our BF16, hidden-dimension-7,168 workload, and the
log transform gives equal spacing to multiplicative changes in payload.
For our eight-Rail topology, $7=R-1$ scales the range from balanced
traffic ($\rho_g=1$) toward the single-Rail extreme. The factor 2 for
$\kappa$ maps the low-contention workload ($\kappa\simeq1$) and the
high-contention workload ($\kappa\simeq3$) to approximately 0 and 1.
These scales are specific to the calibrated configuration and are
re-established when the topology or workload regime changes.

We construct anchors only from repeatable measurements. Both Joint and a
candidate alternative must satisfy $p_{90}/p_{10}\leq1.2$, limiting the
accepted run-to-run spread to 20\%. We further require
$p_{10}(\mathrm{Joint})/p_{90}(\mathrm{candidate})\geq1.05$, so that the
alternative retains at least a 5\% advantage under a conservative
cross-quantile comparison. For low-contention development cases, known
equivalent-path controls are additionally required to agree within 10\%;
otherwise the point is rejected as noisy. These filters leave the four
anchors in Table~\ref{tab:adaptive-anchors}.

\begin{table}[t]
\centering
\small
\caption{Calibrated selector anchors. $B$ is average remote payload per
GPU per All-to-All operation.}
\label{tab:adaptive-anchors}
\begin{tabular}{rrrr}
\toprule
$B$ (MiB) & $\rho_g$ & $\kappa$ & Selected path \\
\midrule
7  & 1 & 1          & Permutation \\
7  & 1 & $\approx3$ & Permutation \\
14 & 1 & 1          & Permutation \\
28 & 1 & 1          & Native \\
\bottomrule
\end{tabular}
\end{table}

At runtime, an anchor is reused only when the payload is within
$1.25\times$ and both normalized shape coordinates are within 0.02.
The payload guard permits limited local reuse while keeping the
7-, 14-, and 28-MiB anchor neighborhoods disjoint. The 0.02 shape
tolerance is deliberately narrow; under the above normalization it
corresponds to deviations of 0.14 in $\rho_g$ and 0.04 in $\kappa$.
Requests outside these neighborhoods fall back to Joint. All thresholds
and anchors were frozen before evaluating the unseen requests below.

\paragraph{Measured selector results.}
We evaluate 24 requests on H800: 12 development-regression requests and
12 requests not used to construct the frozen table. Each path uses five
warmup iterations, followed by interleaved measurements; we report the
median of the first 10 samples. GPU timing covers the complete
dispatch--combine interval and takes the maximum across ranks.

Table~\ref{tab:adaptive-validation} summarizes the results. The selector
switches away from Joint on seven requests, and all seven switches reduce
the measured median latency. On the 12 unseen requests, it switches on
four cases and obtains $1.061$--$1.196\times$ speedup over Joint. Across
all seven switched cases, the measured speedup ranges from
$1.061$--$1.234\times$.

\begin{table}[t]
\centering
\small
\caption{Measured validation of the calibrated selector. Speedup ranges
include only requests where the selector changes the execution path.}
\label{tab:adaptive-validation}
\begin{tabular}{lrrr}
\toprule
Split & Requests & Beneficial switches & Speedup \\
\midrule
Development & 12 & 3/3 & $1.065$--$1.234\times$ \\
Unseen & 12 & {\bfseries\boldmath 4/4} & {\bfseries\boldmath $1.061$--$1.196\times$} \\
\midrule
Total & 24 & {\bfseries\boldmath 7/7} & {\bfseries\boldmath $1.061$--$1.234\times$} \\
\bottomrule
\end{tabular}
\end{table}

The remaining 17 requests fall back to the same Joint implementation.
The Joint and Adaptive columns were nevertheless measured independently,
so differences between them on these requests reflect run-to-run
variation rather than a path-selection effect. We therefore use only the
seven actual path switches to quantify selector benefit.

\subsection{Packing Overhead}
\label{app:packing_overhead}

Permutation-only and Joint require per-wave packing. We compare
\textsc{Prepacked}, \textsc{Sequential}, and \textsc{Pipelined}, where the
last overlaps packing of the next wave with communication of the current wave.
Table~\ref{tab:packing_overhead} reports dispatch--combine median latency.

\begin{table}[t]
\centering
\small
\caption{Packing overhead on H800. Latency is in ms. Pipelined latencies and reductions relative to Sequential are bold.}
\label{tab:packing_overhead}
\begin{tabular}{rrrrr}
\toprule
MiB/GPU & Prepacked & Sequential & Pipelined & Reduction \\
\midrule
10.5 & 3.200 & 3.427 & {\bfseries\boldmath 3.301} & {\bfseries\boldmath 3.68\%} \\
17.5 & 4.747 & 5.127 & {\bfseries\boldmath 4.955} & {\bfseries\boldmath 3.35\%} \\
21 & 5.503 & 5.951 & {\bfseries\boldmath 5.760} & {\bfseries\boldmath 3.21\%} \\
28 & 7.118 & 7.692 & {\bfseries\boldmath 7.447} & {\bfseries\boldmath 3.19\%} \\
35 & 8.642 & 9.372 & {\bfseries\boldmath 9.098} & {\bfseries\boldmath 2.92\%} \\
42 & 10.210 & 11.074 & {\bfseries\boldmath 10.756} & {\bfseries\boldmath 2.87\%} \\
\bottomrule
\end{tabular}
\end{table}

As shown in Table~\ref{tab:packing_overhead}, pipelining consistently reduces
the exposed packing overhead, improving end-to-end latency by
2.87--3.68\% over sequential packing. These differences are measured at the
full communication-path level rather than as isolated packing-kernel times.

\subsection{Static Memory Overhead}
\label{app:memory_overhead}

RailWave trades redundant expert storage for communication workspace.
For $D$ nodes, $G$ local GPUs, per-rank token capacity $T$, and
$C_{\max}=1024$ reserved channels, the dominant RailBalance workspace is
\begin{equation}
M_{\rm RW}\approx
(D-1)(T+C_{\max})(s_d+s_c)
+4(32C_{\max}+2C_{\max}DG+DG),
\end{equation}
where $s_d$ and $s_c$ are the per-record dispatch and return storage.
Under our configuration ($D=4$, $G=8$, $T=50{,}176$, $H=7168$),
$s_d=s_c=2H+32$, giving approximately 4.11~GiB/GPU of additional
workspace. Sequential waves reuse the same workspace rather than allocating
one copy per wave.

EPLB instead incurs persistent storage for redundant experts. If each GPU
keeps $r$ additional experts per MoE layer, with $L$ MoE layers, $P$
parameters per expert shard, and $w$ bytes per weight, its dominant
inference overhead is
\begin{equation}
\Delta M_{\rm EPLB,inf}=LrPw.
\end{equation}
For training, additional replica gradients contribute
$gG_{\rm live}$, where $G_{\rm live}$ depends on how long replica
gradients remain resident.

Table~\ref{tab:memory_overhead} gives a static BF16 example using
DeepSeek-V3-sized routed experts ($L=58$, $P=3\times7168\times2048$)
without expert tensor sharding.

\begin{table}[t]
\centering
\small
\caption{Dominant incremental storage in GiB/GPU. EPLB assumes BF16 expert
weights and FP32 replica gradients; RailWave reports the communication
workspace above.}
\label{tab:memory_overhead}
\begin{tabular}{rrrrr}
\toprule
$r$ & EPLB inf. & Train, 1-layer grad & Train, all grads & RailWave \\
\midrule
1 & 4.758 & 4.922 & 14.273 & {\bfseries\boldmath 4.111} \\
2 & 9.516 & 9.844 & 28.547 & {\bfseries\boldmath 4.111} \\
4 & 19.031 & 19.688 & 57.094 & {\bfseries\boldmath 4.111} \\
\bottomrule
\end{tabular}
\end{table}

\subsection{Replay Workloads and Execution Protocol}
\label{app:replay-workloads}

\paragraph{Motivation.}
Megatron-style training~\citep{shoeybi2020megatron} commonly follows
single-program, multiple-data (SPMD) execution: ranks run the same program
on different data or model partitions and coordinate through communication
and synchronization. This does not require a central controller to track
every rank's fine-grained progress. Compute-time variation, rank skew, and
network jitter can therefore change communication overlap even when the
logical execution order is unchanged. Single-controller runtimes can
explicitly coordinate cross-stage tasks~\citep{xhebraj2024jaxpp}, but
centralized coordination is not a defining property of the multiple-program,
multiple-data (MPMD) model.
We use training-derived layer loads with controlled sending arrangements
to cover possible incast conditions, rather than replaying one incidental
training timeline.

\paragraph{Sending patterns.}
For the four-node replay, each layer's measured 128 expert counts are
sorted and grouped in sets of 16 to obtain eight load bins indexed by
destination-local GPU.
\textsc{Ring} sends from node $s$ to $(s+1)\bmod 4$ in one wave, with
peak incast degree $[1]$. \textsc{Moderate} and \textsc{Extreme} distribute
each source-rank/destination-local-GPU count across the three remote nodes
using integer quotient and remainder, preserving each source's total
volume and destination-local-GPU histogram. They use the same expanded rank-to-rank demand
and three waves, with per-wave peak incast degrees $[2,1,2]$ and $[3,3,2]$, respectively.
Each array entry is the maximum number of source nodes assigned to the
same receiver in that wave. Table~\ref{tab:replay-schedules} specifies the
node-pair assignments. Extreme is a controlled stress case, not an estimate
of how often severe incast occurs in training.

\begin{table}[H]
\centering
\small
\caption{Replay schedules. Each tuple lists the destinations of source nodes
$(0,1,2,3)$; a dash denotes no wave. Incast degree is the scheduled number of concurrent senders to a receiver,
not a measured instantaneous network quantity.}
\label{tab:replay-schedules}
\setlength{\tabcolsep}{5pt}
\begin{tabular}{lcccc}
\toprule
Schedule & Wave 1 & Wave 2 & Wave 3 & Peak incast degree \\
\midrule
Ring        & $(1,2,3,0)$ & --          & --          & $[1]$ \\
Moderate    & $(1,0,0,2)$ & $(2,3,1,0)$ & $(3,2,3,1)$ & $[2,1,2]$ \\
Extreme     & $(1,0,0,0)$ & $(3,3,3,2)$ & $(2,2,1,1)$ & $[3,3,2]$ \\
Permutation & $(1,2,3,0)$ & $(2,3,0,1)$ & $(3,0,1,2)$ & $[1,1,1]$ \\
\bottomrule
\end{tabular}
\end{table}

\paragraph{Backend execution and timing.}
In Moderate and Extreme, Native, Rail-only, and NCCL retain the input
three-wave schedule; Permutation-only and Joint reorder the same demand
into three one-to-one waves. Each remote node pair appears exactly once,
so total volume and wave count are unchanged. Here Native means the
DeepEP backend under the specified replay schedule, not a single call
with the full layer demand. Ring remains a single wave for these paths.
Each sample times all waves of the layer: dispatch--combine per wave for
the DeepEP paths, or forward/return All-to-All for NCCL. CPU planning,
index construction, and prepacking are outside this interval.

\end{document}